# Local 3D real space atomic structure of the simple icosahedral $Ho_{11}Mg_{15}Zn_{74}$ quasicrystal from PDF data

**S. Brühne, E. Uhrig, C. Gross and W. Assmus**

Physikalisches Institut, Johann Wolfgang Goethe-Universität, Robert-Mayer-Str. 2-4,
D-60054 Frankfurt/Main, Germany

We present a new complementary strategy to quasicrystalline structure determination: The local atomic structure of simple icosahedral (*si*) $Ho_{11}Mg_{15}Zn_{74}$ [$a(6D) = 5.144(3)$Å] in a sphere of up to $r = 17$Å was refined using the atomic pair distribution function (PDF) from in-house X-ray powder diffraction data (MoK$\alpha_1$, $Q_{max} = 13.5$Å$^{-1}$; $R = 20.4\%$). The basic building block is a 105-atom Bergman-Cluster {$Ho_8Mg_{12}Zn_{85}$}. Its center is occupied by a Zn atom – in contrast to a void in face centred icosahedral (*fci*) $Ho_9Mg_{26}Zn_{65}$. The center is then surrounded by another 12 Zn atoms, forming an icosahedron (1st shell). The 2nd shell is made up of 8 Ho atoms arranged on the vertices of a cube which in turn is completed to a pentagon dodecahedron by 12 Mg atoms, the dodecahedron then being capped by 12 Zn atoms. The 3rd shell is a distorted soccer ball of 60 Zn atoms, reflecting the higher Zn content of the *si* phase compared to the *fci* phase. In our model, 7% of all atoms are situated in between the clusters. The model corresponds to a hypothetical 1/1-approximant of the icosahedral (*i*) phase. The local coordinations of the single atoms are of a much distorted Frank-Kasper type and call to mind those present in 0/1-$Mg_2Zn_{11}$.

## 1   Introduction

In contrast to the atomic structures of periodic crystals, the structures of non-crystalline solids, such as amorphous materials or glasses, are only ambiguous when detecting and describing them. Although, at first glance they suffer from an overall statistical disorder, their local atomic environments or average coordination shells are accessible from powder diffraction. Thus, their one-dimensional (1D) diffractogram $I(Q)$ can be transformed from reciprocal space to its 1D real space counterpart $G(r)$. The latter is the radial atomic pair distribution function (PDF) which represents the probability for finding any atom i (*i.e.* in the case of X-rays an electron density maximum) in the distance $r_{ij}$ to any other atom j with respect to the average electron density of the sample. Recent examples for quantitative investigations of such materials are vitreous $V_2O_5$ [1] or $Ca_{x/2}Al_xSi_{(1-x)}O_2$ glasses [2].

In addition, disorder, which is contained within crystalline material, is readily quantified by analysis of the PDF (the PDF includes *non*-Bragg diffuse scattering and therefore contains information on the disorder): *E.g.* the local structure of the solid solution $Ga_{(1-x)}In_xAs$ was determined in [3]. Exfoliated-restacked *r*-$WS_2$ that was considered to be amorphous in the first place, could be shown to obtain monoclinic translational symmetry (Pearson Symbol [4]: mP12) [5].

Already in the late 1980´s, in the search for similarities to related phases of quasicrystals, pair distribution functions of *e.g.* icosahedral (*i*) Pd-Si-U have been compared to those of amorphous *A*-Pd-Si-U [6] or decagonal (*d*) Al-Mn to α-Al-Mn-Si (cP138) [7]. Among other many different approaches for gaining an insight into the atomic structure of quasicrystals – such as the tiling models from HRTEM images (*e.g.* for *d*-Al-Co-Ni [8]), X-ray holography (*i*-Al-Mn-Pd [9]), EXAFS measurements (*e.g. i*-Dy-Mg-Zn [10]) or pair potential simulations (*e.g.* for *i*-Al-Mn [11]) – the higher dimensional (*n*D; $n > 3$) crystallography yields the most accurate information [12]. Single crystal data sets permit a quite high ratio *N*(data)/*N*(variables) from and the *n*D models describe completely the quasiperiodicity. Nevertheless, its drawbacks are the high experimental, formalistic and computational input – it is also not intuitive in deriving the 3D parallel (or physical) real space meaning of the hyperatoms in the *n*D structure model.

As a complementary and novel route to the 3D real space atomic structure of quasicrystals, the PDF can be used to refine local models with a least squares algorithm: This was achieved in our laboratory recently to evaluate quantitatively the local and medium range (in a sphere of $r = 27$Å) structure of a ternary alloy of composition $Ho_9Mg_{26}Zn_{65}$ [13]. It has a face centered icosahedral (*fci*) 6D lattice ($a$(6D) = 2 × 5.18(3)Å, $F2/m$-3-5). From our results, it can be shown in 3D to be composed of 3-shell, 104-atom Bergman clusters ($d \approx 15$Å) comprising a void at the center. The clusters are interconnected sharing common edges and hexagonal faces of the 3rd shells. The remaining space is filled by some glue atoms, yielding an almost tetrahedrally close packed structure. All Ho atoms are situated in the 16-atom Frank-Kasper- (FK-) polyhedron "P", most of them being situated in the 2nd shell. It would seem then that the long range structure of *fci*-Ho-Mg-Zn can be anticipated as a quasiperiodic canonical cell tiling (CCT, after Henely [14]) decorated with Bergman clusters.

In the ternary Ho-Mg-Zn system we have found another *i* phase, more rich in Zn, but with a simple icosahedral (*si*) 6D lattice: *si*-$Ho_{11}Mg_{15}Zn_{74}$, $a$(6D) = 5.143(2)Å [15]. "Simple" denotes the crystallographically *primitive* 6D Bravais type. The possible superspace groups are $P2/m$-3-5 or $P235$ according to reflection conditions [16]. Single crystal diffractograms show virtually no diffuse scattering pointing to high structural perfection. Thus, pursuant to [17], this *si*-phase could be a promising candidate for quasiperiodic order of the magnetic moments carried by the 4*f* electrons localized at the Ho atoms. While the considerations in [17] deal with the general case – it is still desirable to know the arrangement of Ho in 3D space for the title compound. Hence we tackled the structure of *si*-Ho-Mg-Zn analogously to the *fci*-phase. This paper presents our first results.

After a general explanation of our new method for local structure determinaton in Sec. 2, the synthesis, characterisation and structure analysis of *si*-Ho-Mg-Zn will be described in Sec. 3. The resulting data will be discussed, mostly in terms of crystal chemistry in Sec. 4; Sec. 5 will conclude the paper.

## 2     Method

Once material is available in a single phase and good (quasi)crystalline quality, our strategy is as follows:
(i)     record X-ray powder diffraction data $I(Q)$ and generate $S(Q)$
(ii)    generate the atomic pair distribution function $G(r)$ (PDF)
(iii)   build local structure model(s)
(iv)    verify and refine the structure model by fitting the PDF data

*Ad* (i): As a good resolution of experimental PDFs in $r$ mainly arises from a large $Q_{max} = 4\pi\sin(\theta_{max})/\lambda$, radiation of preferably short wave length is recommended and measurement accurate in high $Q$ has to be performed [18]. The software PDFgetX [19] was used for data reduction to extract the elastic part of the total diffracted intensity $I^{el}(Q)$, and to calculate the total structure function $S(Q)$ according to equation (1). $c_i$ are the atomic concentrations (from WDX analysis; see Sec. 3) and $f_i$ the scattering factors for the atomic species i.

$$S(Q) \;=\; 1 + \{\, I^{el}(Q) - \Sigma_i\, c_i\, [f_i(Q)]^2 \,\} / \{\, \Sigma_i\, c_i\, f_i(Q) \,\}^2 \qquad (1)$$

*Ad* (ii): The PDF from X-rays is defined in equation (2) where $\rho$ is the electron density in the sample, $\rho_0$ being its average value. The experimental PDF is accessible *via* sine Fourier transformation of $S(Q)$ according to equation (3). This is done using PDFgetX [19] again.

$$G(r) \;=\; 4\pi r\, [\rho(r) - \rho_0] \qquad (2)$$

$$G(r)_{exp} \;=\; 2/\pi \int_0^\infty Q[S(Q) - 1]\,\sin(Q\,r)\,dQ \qquad (3)$$

*Ad* (iii): A theoretical PDF of a model structure can be calculated from equation (4), $\langle f \rangle$ is the average scattering factor of the model:

$$G(r)_{calc} \;=\; 1/r\, \Sigma_i \Sigma_j \left[\, f_i f_j/\langle f\rangle^2\; \delta(r - r_{ij})\,\right] - 4\pi r\, \rho_0 \qquad (4)$$

To obtain *local* model structures of *i* quasicrystals, the concept of 3D *rational* approximants for the 6D hypercrystals can be used: They are based on the *irrational* cut-and-projection technique to generate a 3D quasiperiodic structure from the 6D periodic structure. If the slope of 3D space *is not* irrational (*i.e.* including the "golden mean" $\tau = (\sqrt{5} + 1)/2$) as the generic factor) with respect to 6D space, but a *rational* (Fibonacci) ratio $p/q$ instead, such periodic approximants result. In the icosahedral case, their point symmetry has to be a subgroup of $m$-3-5, most convenient are cubic structures with Laue symmetry $m$-3. In equation (5) their 3D lattice parameter $a(3D)$ is related to the 6D hypercubic lattice constant $a(6D)$:

$$a(3D) \;=\; 2\, a(6D)\, (p\tau + q) / \sqrt{2 + \tau} \qquad (5)$$

Other crystal systems besides cubic are possible, the cells can be decorated by icosahedral clusters obeying the rules for a canonical cell tiling (CCT), see [14]. *E.g.* in the case of Al-Mg-Zn and Ga-Mg-Zn, a series of rational approximants could be prepared as their structures were already determined conventionally – the CCT playing a major rôle [20, 22] (Tab. 1). The CCT is built up of four sorts of canonical cells, namely A, B, C and D (compare Fig. 7). Though lower approximants can be constructed today [21], the *long range* quasiperiodic tiling is still waiting to be found [23]. This quasiperiodic tiling is highly desirable because together with a decoration rule it would be an outright description for an *i* quasicrystal.

| *p/q* | space group | *a*(3D)/Å | Pearson symbol | example | CCT cell content [14] | ref. |
|---|---|---|---|---|---|---|
| 1/0 | *Pm*-3 | 8.6 | cP39 | Mg$_2$Zn$_{11}$ | – | [24] |
| 1/1 | *Im*-3 | 14 | cI160 | Al-Mg-Zn | 12A | [22] |
| 2/1 | *Pa*-3 | 23 | cP680 | Al-Mg-Zn | 24A 8B 8C | [22] |
| 3/2-2/1-2/1 | *Cmc*2$_1$ | 37-23-23 | oC1108 | Ga-Mg-Zn | 24A 12B 12C 4D | [20] |
| 3/2 | *Pa*-3 | 37 | cP2888 | – | 72A 32B 32C 8D | [20] |
| 5/3 | *P*2$_1$3 | 60 | cP~12200 | – | - ? - | [25] |

**Tab. 1** Rational approximants of *i*-Mg-(Zn,*B*) (*B* = Al, Ga) as bases for approximant models of *i*-Mg-Zn-*RE* (*RE* = Y and some rare earths)

Actually, in the Mg-Zn-*RE* systems such rational approximants have *not* been observed experimentally so far. Nevertheless, rational "1/1" and "2/1" approximant *models* of *fci*-Ho-Mg-Zn have been successfully used by us to refine and describe its *local* structure [13]. For the Ni-Ti-Zr system, recently a hypothetical "8/5" approximant (*a'*(3D) = 98Å) has been constructed to describe *i*-Ni-Ti-Zr [26].

*Ad* (iv): If such a "*p/q*"-approximant model can be set up, its calculated PDF can be fitted to the experimental PDF of the *i* phase by refining the model. As a coordinate system, a hypothetical cubic lattice with a period *a*´(3D) can be used. This corresponds to *periodic* boundary conditions on the local model of the intrinsic *aperiodic* structure, therefore the radius of the model PDF is (arbitrarily) confined to $r_{max} \approx 1.2 \times a'(3D)$. Technically, the refinement procedure is comfortably done using PDFFIT, a least-squares refinement program [27] which is handled like a Rietveld refinement. The quantitiy minimized is *R* as defined in equation (6); i: data points.

$$R = \sqrt{\{ \sum_{i=1}^{N} [G_{obs}(r_i) - G_{calc}(r_i)]^2 / \sum_{i=1}^{N} G^2_{obs}(r_i) \}} \qquad (6)$$

In contrast to a Rietveld refinement, the only "profile" parameter fitting the PDF is the dynamic correlation factor δ which accounts for correlated thermal motion [28]. Reasonable *R* values range from 0.10 to 0.35, depending on a variety of factors such as the resolution of the PDF (equivalent to $Q_{max}$), *r* range, the numbers *N*(data points) or *N*(parameters), and therefore have to be taken *cum grano salis* [5, 13]. After final convergence, a detailed analysis regarding the chemical and physical soundness of the resulting structure should always be carried out.

## 3     Experimental

*Synthesis*

Based on our results in [15], the LETSSG method [29] was used to grow large single crystals of the title compound (Figure 1):
100 g of the pure elements zinc (99.9999%), magnesium (99.99%) and holmium (99.9%) were placed into an aluminia crucible in a ratio $Zn_{62.8}Mg_{33.6}Ho_{3.6}$. 20g of an eutectic LiCl/KCl mixture is added to form an encapsulation of the melt when heated. This is necessary to prevent the zinc from evaporating during crystal growth due to the high vapour pressure $p(Zn)$ at the required temperature. The crucible is heated under argon atmosphere up to 800°C where Zn and Mg are molten and the Ho dissolves in the Mg-Zn melt. After a homogenisation time, a water-cooled tungsten tip is placed into the melt and the temperature of the furnace is lowered at a rate of 0.4 K/h. When the temperature reaches 720°C, the tungsten tip with the crystal is pulled from the melt. For characterization purposes, parts of the triacontahedrally facetted single crystal were cut off by spark erosion.

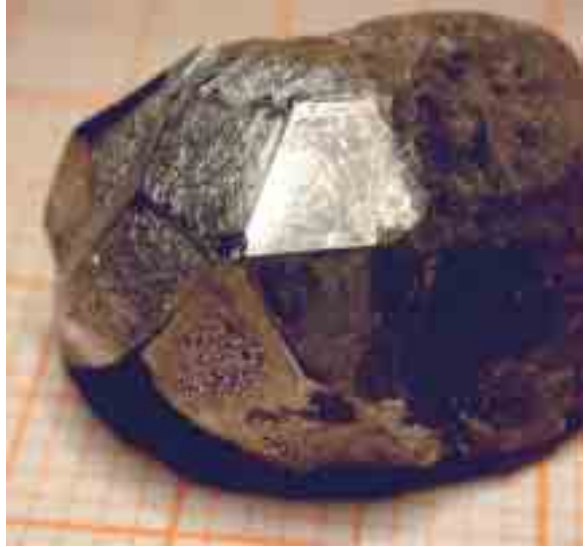

**Fig. 1**  *si*-Ho-Mg-Zn: Centimeter sized single crystal of triacontahedral morphology

*Characterization*

The obtained crystal has a mass density of 6.46±0.02 g/cm$^3$, measured by a gas-pycnometer „Micromeritics AccuPyc 1330". The composition determined by WDX (Microspec WDX3PC is $Ho_{11}Mg_{15}Zn_{74}$: Polished samples were measured *versus* standard specimen of the pure metals with an accuracy of ± 1 at.%. 12mg of the sample was submitted to differential dissolution technique [30, 31]: Ion coupled plasma (ICP) elemental analysis revealed homogenity and confirmes the WDX result. A possible presence of Li in the quasicrystal can clearly be ruled out. Its icosahedral symmetry was confirmed by X-ray Laue photographs. A portion was crushed and its X-ray powder diffractogram (Siemens Kristalloflex 810, CuK$\alpha$, $\lambda$ = 1.541Å) could be indexed with a *si* lattice parameter $a(6D)$ = 5.144(3)Å which coincides with that found in [15].

Diffractograms of another portion were measured on a Huber Guinier Diffractometer (Seifert system 600) using MoK$\alpha_1$ radiation ($\lambda$ = 0.70932Å; $2\theta$ = 4.. 100°; $\Delta\theta$ = 0.01°; $t$ = 60s). High angle data ($2\theta > 60°$) were measured in a second run with triple measuring time. The curves were averaged and background corrected by subtraction of a spline function. Applying the computer program PDFgetX, the data were corrected for multiple scattering, polarisation and absorption effects. The structure function $S(Q)$ (Figure 2) was obtained by normalisation and finally the experimental PDF $G(r)_{exp}$ (equation 1) was calculated [19].

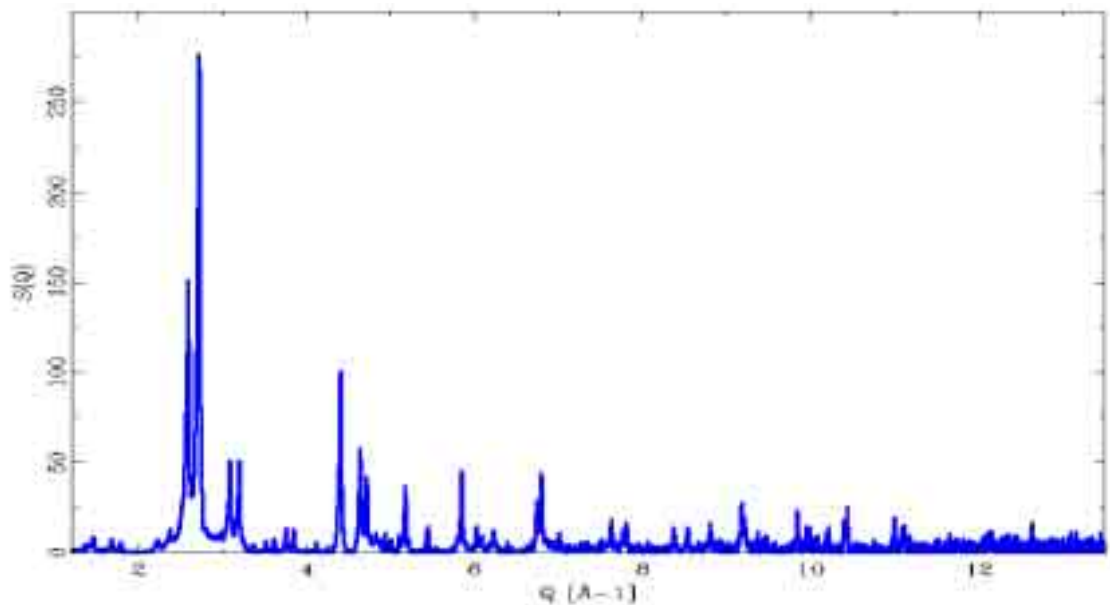

**Fig. 2** Structure function $S(Q)$ of $si$-Ho$_{11}$Mg$_{15}$Zn$_{74}$ from corrected and normalized MoK$\alpha_1$ diffraction data ($2\theta_{max} = 100°$).

*Local structure determination*

A crucial step in our method is to find a reasonable model from which to start the PDF refinements. Luckily, in the case of $si$-Ho-Mg-Zn, the similarity of its diffractogram to $fci$-Ho-Mg-Zn [15] points to very similar structural features, as does its PDF. Thus, the model of "1/1"-Ho$_{12.0}$Mg$_{28.0}$Zn$_{60.0}$ for the $fci$ phase [13] was used as a starting point. An initial hypothetical lattice constant was set to $a'(3D) = 14.0$Å [from $a(6D) = 5.14$Å in eq. (5)]. As maximum refinement range, arbitrarily $r_{max} = 17$Å $\approx 1.2 \times a'(3D)$ was chosen. The symmetry constraints of space group $Im$-3 were applied. The refinements benefit from the good contrast for X-rays of Mg, Zn and Ho ($Z = 12$, 30 and 67), respectively.

Refinement of the scale factor, $a'(3D)$, and the dynamic correlation factor $\delta$ yield $R = 45\%$ in a range $r = 2 .. 17$ Å. Including the (isotropic) temperature factors $U_{eq}$ for the atoms, $R$ was then lowered to 38% and $U_{eq}$(Mg3) and $U_{eq}$(Mg4) of the $fci$ model approached zero value. To account for the missing electron density here, they were replaced by zinc atoms, labelled Zn4 and Zn5, respectively (see Tab. 3). This already yields a reasonable model composition Ho$_{10.0}$Mg$_{15.0}$Zn$_{75.0}$, much akin to the above WDX value. Also, the positional parameters were allowed to refine at this point and convergence at $R = 21.9\%$ was achieved. A first analysis of the structure revealed reasonable local coordinations, yet were much distorted when compared to the $fci$ phase. The diameter of the first cluster shell offers space for an extra Zn atom at the cluster center. Indeed, after insertion of Zn0 at 000 the $R$ value again decreases to 20.4%. This end point is slightly better than for our "1/1"- model of the $fci$ phase (21.8%, [13]).

To check whether the distribution of Ho and Mg atoms about the β positions in the 2nd shell is like the "cube arrangment" in the $fci$ phase [13] or tends to be random, the occupancies of Mg1 and Ho1 were refined. Again, convergence to $R = 20.5\%$ was reached, while o[Mg1] = 1.08(1) and o[Ho1] = 1.007(1), *i.e.* virtually equal to one. This "cube" feature should then be present in both the *fci* and *si* phases.

Figure 3 compares measured, calculated and difference PDFs. The refined data were used to calculate bond lengths and coordination numbers and were employed for the drawings of the structure (DIAMOND, [32]). Table 2 comprises the data of the refinement. Note that using eq. (5) the 6D lattice parameter of the quasicrystal can be re-determined from the PDF data to $a(6D) = 5.119$Å.

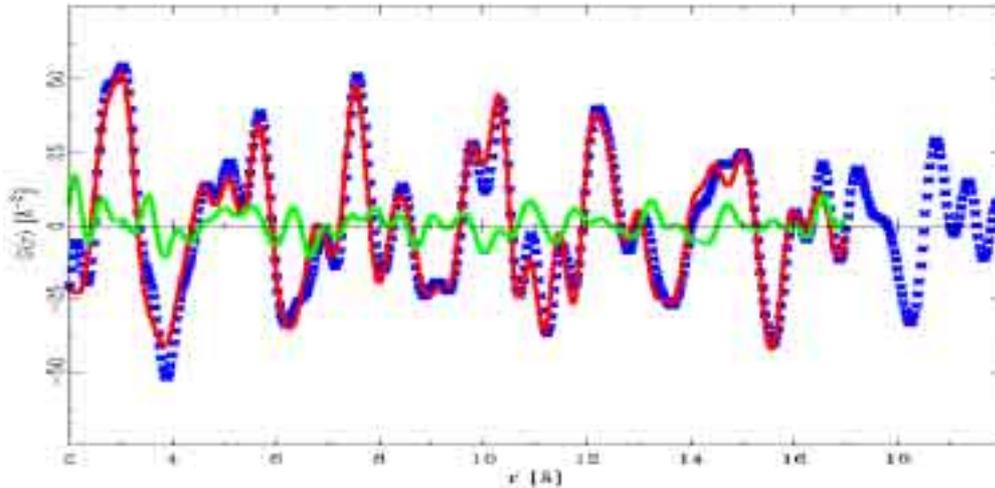

**Fig. 3** PDF from measured diffraction data of *si*- $Ho_{11}Mg_{15}Zn_{74}$ (cross), PDF calculated from the approximant model "1/1"-$Ho_{9.9}Mg_{14.8}Zn_{75.3}$ ($r_{max}$ = 17Å, black line) and their difference plot; $R = 20.4\%$ (grey line).

| refinement data | "1/1"-$Ho_{9.9}Mg_{14.8}Zn_{75.3}$ |
|---|---|
| scale factor | 15.20(3) |
| dynamic correlation factor δ | 0.337(11) |
| low $r/\sigma$ ratio | 1.0 |
| hypothetical approximant space group [33] | *Im*-3 (no. 202) |
| hypothetical approximant lattice parameter $a´$ (3D) /Å | 14.0924(5) |
| hypercubic lattice parameter $a(6D)$ /Å (from eq. 5) | 5.119 |
| data range in $r$ /Å | 0.. 30 |
| calculated $r$ range in Å | 0.. 19.963 |
| refinement $r$ range in Å | 2.. 17 |
| number of data points used | 499 |
| radiation | MoK$\alpha_1$ |
| λ/Å | 0.70932 |
| termination at $Q_{max}$ /Å$^{-1}$ | 13.5 |
| $Q$ resolution σ($Q$) /Å$^{-1}$ | 0.01 |
| number of refined parameters | 24 |
| magic number (relaxation factor) | 20.0 |
| *R*-values | 0.20440322 |
| change last cycle | -0.00179814 |
| correlations greater than 0.8 | none |

**Tab. 2** Refinement data for the "1/1" model of *si*- $Ho_{11}Mg_{15}Zn_{74}$

# 4 Results and Discussion

Table 3 summarizes the structural parameters of our refined local model: Eight atomic orbits are symmetry related *as if* in space group *Im*-3 [33] with a hypothetical translational period of $a'$(3D) = 14.09Å. The composition $Ho_{9.9}Mg_{14.8}Zn_{75.3}$ is derived from the data fitting the WDX results quite well. The density is calculated to $\rho_{X\text{-ray}}$ = 6.65gcm$^{-3}$ (as compared with the measured value of $\rho_{meas}$ = 6.46(2)gcm$^{-3}$, Sec. 3). The temperature factors $U_{eq}$ range from $0.5\times10^{-2}$Å$^{-2}$ to $3.1\times10^{-2}$Å$^{-2}$. They presumably "bury" the difference between our local structure model and the real local structure of the quasicrystal. $d_{min}$, $d_{max}$ and $<d>$ indicate local interatomic distances and are credibly realistic, as are the coordination numbers and geometries. To faciliate comparisons to other (crystalline) intermetallic alloys, the geometries are described by the polyhedron code defined in [34], as it denotes vertex configurations *j* in terms of polygons meeting in one vertex: $N_1^{triangles.\ quadrangles.\ pentagons...}$ $N_2^{...}$ ... $N_3$ ... and $\Sigma N_j$ = CN; *e.g.* tetrahedron: $4^{3.0}$; cube: $8^{0.3}$; octahedron: $6^{4.0}$. The Frank Kasper (FK) codes X, P, Q and R will be explained below. The last column assigns structural functions to the distinct atomic orbits following Kreiner's nomenclature [20]: Atoms with CN12 are labelled α, CN16 - β, CN14 - γ and CN15 - δ. An additional superscript specifies the structural function, see the following paragraph "Basic structural unit".

| atom | x | y | z | $U_{eq}$/ $10^{-2}$ Å$^2$ | multi-plicity | $d_{min}$/Å | $d_{max}$/Å | $<d>$/Å | CN | polyhedron code [34] (FK) | structural function |
|---|---|---|---|---|---|---|---|---|---|---|---|
| Zn1 | 0.0870 | 0.3060 | 0.3394 | 0.8 | 48 | 2.452 | 3.471 | 2.903 | 12 | $12^{5.0}$ (X) | $α^3$ - soccer ball |
| Mg1 | 0 | 0.119 | 0.268 | 0.8 | 24 | 2.432 | 4.110 | 3.202 | 16 | $12^{5.0}4^{6.0}$ (*) | β - pentagon dodecahedron |
| Zn2 | 0 | 0.1603 | 0.0994 | 0.5 | 24 | 2.432 | 3.031 | 2.745 | 12 | $12^{5.0}$ (X) | $α^1$ - inner icosahedron |
| Zn3 | 0 | 0.3251 | 0.1681 | 2.7 | 24 | 2.516 | 3.308 | 2.900 | 12 | $12^{5.0}$ (X) | $α^2$ - outer icosahedron |
| Ho1 | 0.1917 | *x* | *x* | 3.1 | 16 | 2.846 | 3.308 | 3.100 | 16 | $12^{5.0}4^{6.0}$ (P) | β - pentagon dodecahedron/ cube |
| Zn4 | 0.1289 | 0 | ½ | 3.0 | 12 | 2.526 | 4.110 | 3.216 | 14 | $12^{5.0}2^{6.0}$ (*) | γ - soccer ball |
| Zn5 | 0.3168 | 0 | ½ | 0.6 | 12 | 2.648 | 3.406 | 3.114 | 15 | $12^{5.0}3^{6.0}$ (Q) | δ - glue atom |
| Zn0 | 0 | 0 | 0 | 1.6 | 2 | 2.658 | 2.658 | 2.658 | 12 | $12^{5.0}$ (X) | $α^0$ - cluster center |

**Tab. 3** Structural parameters of *si*-$Ho_{11}Mg_{15}Zn_{74}$ refined as an approximant model "1/1"-$Ho_{9.9}Mg_{14.8}Zn_{75.3}$ *as if* in *Im*-3 (R = 20.4%). *x*, *y* and *z* are fractions of $a'$ = 14.09Å; $d_{min}$ ($d_{max}$; $<d>$) is the minimal (maximal; average) distance of an atom from its coordinating neighbours. (*): no FK polyhedron is assigned.

*Local coordinations*

Figure 4 shows the local coordination polyhedra of all atomic orbits. While Zn0 is co-ordinated by 12 Zn2 atoms with almost ideal icosahedral symmetry (a feature known from 0/1-$Mg_2Zn_{11}$ [24]), the other atoms reside in more or less distorted coordination geometries. Note *e.g.* the evident off-center position of Zn4. After detailed analysis using [32], a polyhedron code [34] can be assigned to each (Tab. 3). Additionally, X, P, Q and R mark FK triangulated coordination shells of CN 12, 16, 15 and 14, respectively, as defined in [35, 36]. However, the distortion of Mg1 and Zn4 coordination shells is very large – therefore, in the strict sense of [36, 37], one cannot speak of a FK or tetrahedrally close packed (*tcp*) type structure any more. Thus no FK code is assigned for those atoms.

All CN 12 positions are occupied by Zn atoms; the environments of Zn4 (CN 14) and Zn5 (CN 15) would also allow for substitution by Mg atoms. Interestingly, for the solidification sequence of *si*-Ho-Mg-Zn a phase width $Ho_{10}Mg_{14+\delta}Zn_{76-\delta}$, $\delta = 0..6$ was observed in [15]. This corresponds to the Zn4 and Zn5 sites offering space for substituting Mg atoms.

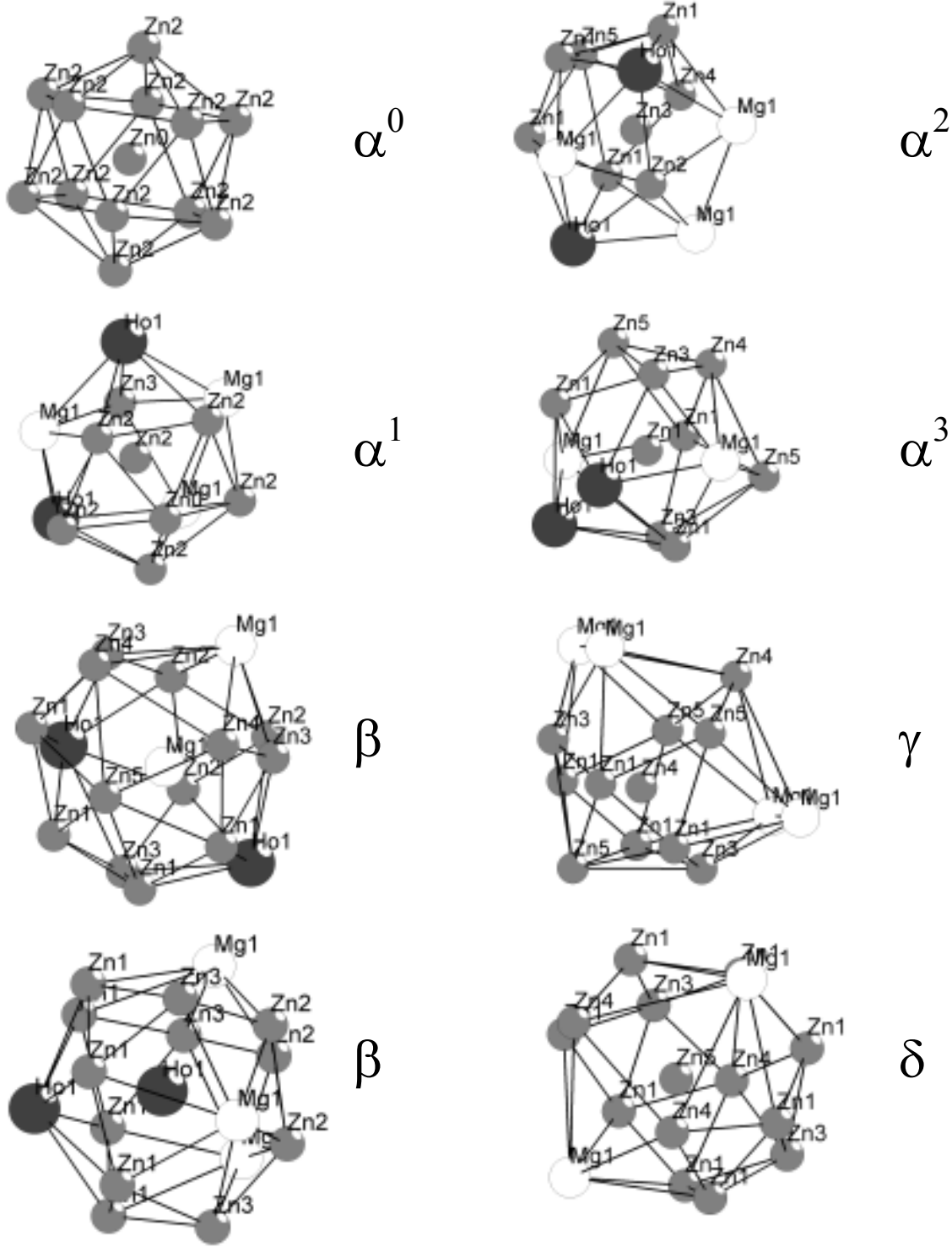

**Fig. 4** Local coordination polyhedra in the refined "1/1"-$Ho_{9.9}Mg_{14.8}Zn_{75.3}$ model for *si*-$Ho_{11}Mg_{15}Zn_{74}$, labelling: see text.

Figure 5 plots the average nearest neighbour distance $<d>$ *versus* the coordination number CN. The points are much scattered in the plot, reflecting the distortions mentioned above. However, a general tendency towards larger mean distances $<d>$ for higher CNs can be observed and substantiates the structure model. A similar scattering for coordination geometries can be found for the neighbouring phase 0/1-$Mg_2Zn_{11}$ (cP39, *Pm*-3 [24]). If one analogously compares *fci*-$Ho_9Mg_{26}Zn_{65}$ to its neighbouring Frank Kasper phase $MgZn_2$ (Laves phase, hP12, $P6_3/mmc$ [38]) as done in [13], more regularity (as expected for *tcp* structures) is present. To achieve good *tcp* packings, an optimum ratio of numbers of smaller ($d_{av}(Zn)$ = 2.788Å [39]) and larger atoms ($d_{av}(Ho)$ = 3.532Å and $d_{av}(Mg)$ = 3.204Å [39]) has to be met.

Thus, as a result, the higher Zn content of *si*-$Ho_{11}Mg_{15}Zn_{74}$ and $Mg_2Zn_{11}$, compared to *fci*-$Ho_9Mg_{26}Zn_{65}$ and $MgZn_2$, causes more distortion of the local coordination shells.

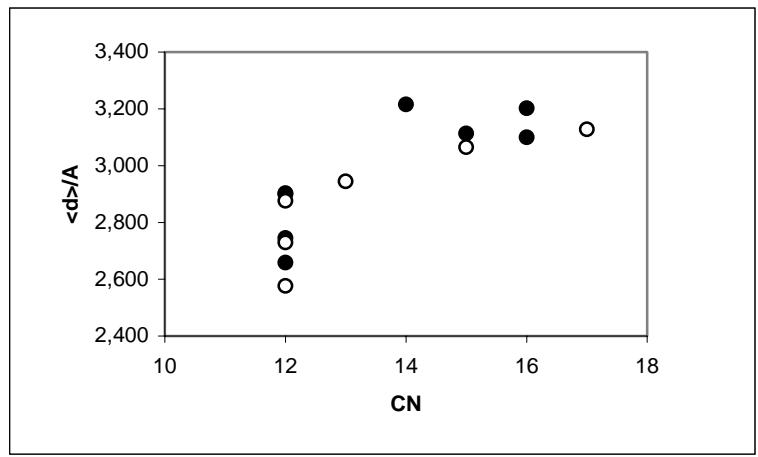

**Fig. 5** Average coordinating distances $<d>$ are increasing with the coordination number CN; filled circles: "1/1" model for *si*-$Ho_{11}Mg_{15}Zn_{74}$, open circles: $Mg_2Zn_{11}$ for comparison

Table 4 compiles the data of average local surroundings of the different atomic species in our "1/1" model. They may *e.g.* serve for a comparison to data from future EXAFS investigations of *si*-$Ho_{11}Mg_{15}Zn_{74}$. The CNs correspond reasonably to the mean intermetallic atomic diameters of the different metals (see above; [39]). An exception to this is the Ho-Ho distance: $d$ = 2.85Å is too small compared to $d_{av}(Ho)$ = 3.532Å. As will be discussed below, it can be speculated that no such direct Ho-Ho contacts are present in the real quasiperiodic structure. Besides, there is only one Ho-Ho-contact per average Ho atom in the present model.

| "average" atom i | $N_j$(neighbours j) ( $<d_{i,j}>$/Å ) | | | $\Sigma N_j$ = |
|---|---|---|---|---|
| | **Ho** | **Mg** | **Zn** | **<CN>** |
| **Ho** | 1.0 (2.85) | 3.0 (3.08) | 12.0 (3.13) | 16.0 |
| **Mg** | 2.0 (3.08) | 1.0 (3.35) | 13.0 (3.20) | 16.0 |
| **Zn** | 1.6 (3.13) | 2.6 (3.20) | 8.4 (2.79) | 12.6 |

**Tab. 4** Average data for the first coordination shells in *si*-$Ho_{11}Mg_{15}Zn_{74}$: average numbers $N_j$ of nearest neighbours j for average atom i and average coordination numbers <CN> in the "1/1"-$Ho_{9.9}Mg_{14.8}Zn_{75.3}$ model

*Basic structural unit*

Further analysis reveals the (approximately icosahedral) principal building block of the structure, see Fig. 6. It can be described as a 3-shell 105-atom Bergman cluster [40]:
A Zn atom in its center ($\alpha^0$) is surrounded by 12 further Zn atoms on the vertices of a regular (inner) icosahedron ($\alpha^1$, $r = 2.7$Å). This icosahedron is identical to the first coordination shell of Zn0 (Fig. 4). A second shell is found from $r = 4..5.5$Å: A Ho$_8$-cube of edge length $a = 5.4$Å ($\beta$, $r = 4.7$Å) is completed by 12 Mg atoms ($\beta$, $r = 4.1$Å) to an irregular pentagon dodecahedron. Another 12 Zn atoms ($\alpha^2$, outer icosahedron) are assigned to this shell because of its similar radius of $r = 5.1$Å. A third shell, made up of 48 $\alpha^3$ ($r = 6.8$Å) and 12 $\gamma$ atoms ($r = 7.2$Å) consists of Zn atoms only. This 60-atom shell has the shape of a distorted soccer ball.

The clusters are linked *via* shared hexagonal faces of the soccer ball (8 faces per ball) and common edges ($\gamma$ atoms; 6 edges per soccer ball). Some $\delta$ atoms act as glue or fillers inbetween the clusters. Thus one can write the formula $\{[\alpha^0\alpha^1{}_{12}(\beta_{8+12}\alpha^2{}_{12})(\alpha^3{}_{48-8\times6/2}\gamma_{12/2})]\delta_6\}$ or $\{[ZnZn_{12}(Ho_8Mg_{12}Zn_{12})(Zn_{48-8\times6/2}Zn_{12/2})]Zn_6\}$ for the whole structure in terms of the basic cluster.

It should be noted that the term "cluster" may be misleading sometimes: In the structure discussed here no separation in terms of different bonding principles (as different as *e.g.* in Chevrel phases [41]) is observed. One can see from Fig. 4 that the local environments of $\alpha^3$ and $\gamma$ atoms in the outer cluster shell or the $\delta$ glue atom *inter* cluster in principle are not different from the environments of *intra* cluster atoms. Consequently, though the structure is not exactly *tcp*, it is an almost isotropic packing of metal atoms.

*Cluster connecting scheme*

From the sharing of faces and edges of the third shell as explained above, a body centred cubic (*bcc*) arrangement of the clusters arises. As depicted in Figure 7b, it can be completely decomposed into A cells of Henleys CCT [14]. According to [14] the A cell has two edges of length $b = a'(3D) \approx 14$Å (so-called b-bond) and four c-bonds of length $c = \sqrt{3}/2\, b \approx 12$Å. b-bonds run along the edge sharing (2-fold) connections of the soccer balls, a c-bond runs along the 3-fold hexagon sharing connection. Figure 7b corresponds to a 1/1 approximant; for higher approximants additional CCT cells B, C and D are required (Fig. 7a). Our model thus represents the first step towards the construction of a quasiperiodic decorated CCT: In the future we look forward to the construction and refinement of higher approximant models, then finally for a quasiperiodic tiling model of the *si*-Ho$_{11}$Mg$_{15}$Zn$_{74}$ phase. The next higher ("2/1") model has already been refined for the *fci* phase [13]. CCT node environments and therefore the cluster connecting scheme are slightly different then, though the first two shells of the clusters are left invariant. We believe that this will also hold for the *si* phase.

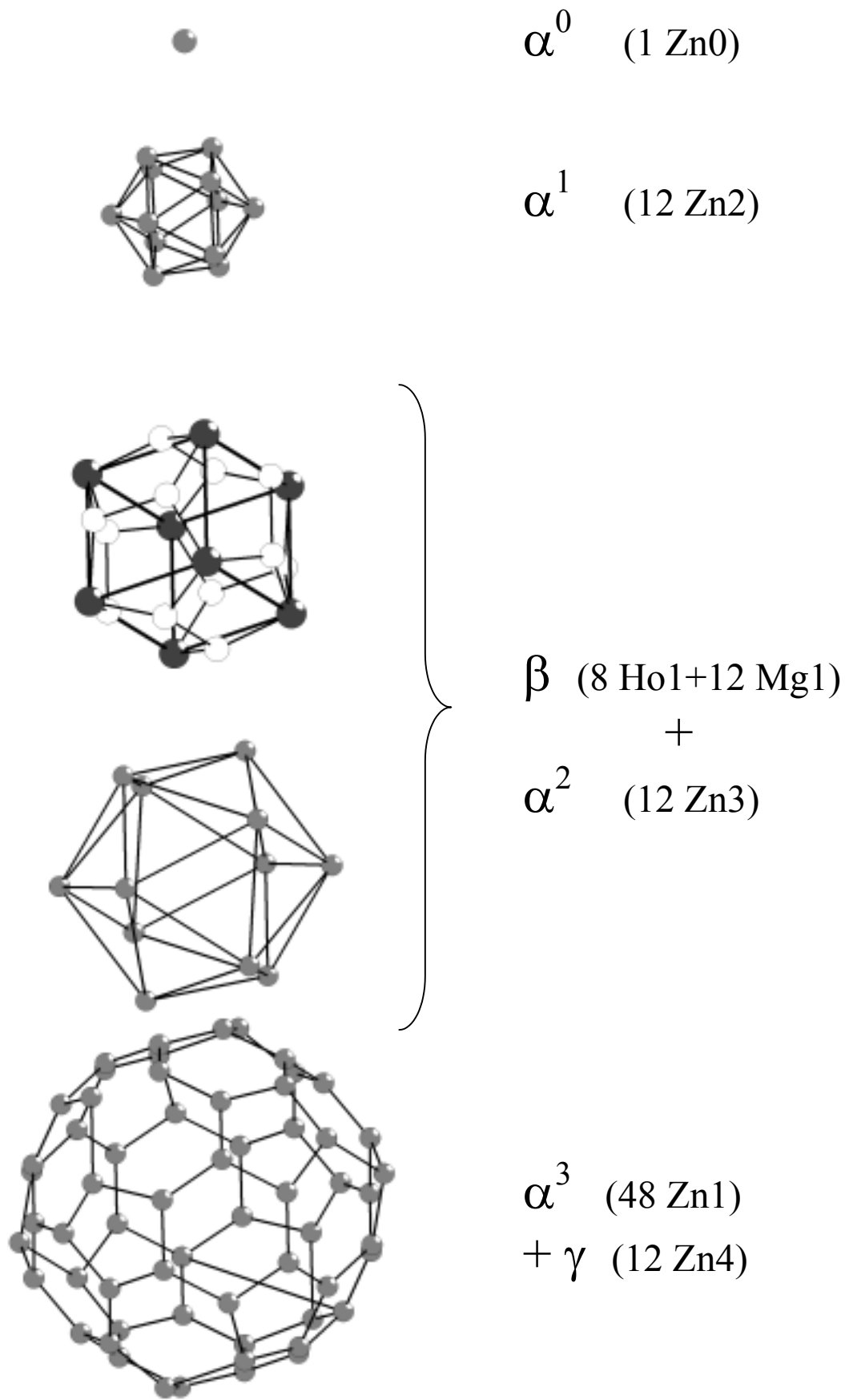

**Fig. 6** 3-shell 105-atom Bergman cluster in the "1/1"-$Ho_{9.9}Mg_{14.8}Zn_{75.3}$ model for *si*-$Ho_{11}Mg_{15}Zn_{74}$. Mg atoms: white, Zn atoms: grey and Ho atoms: dark grey.

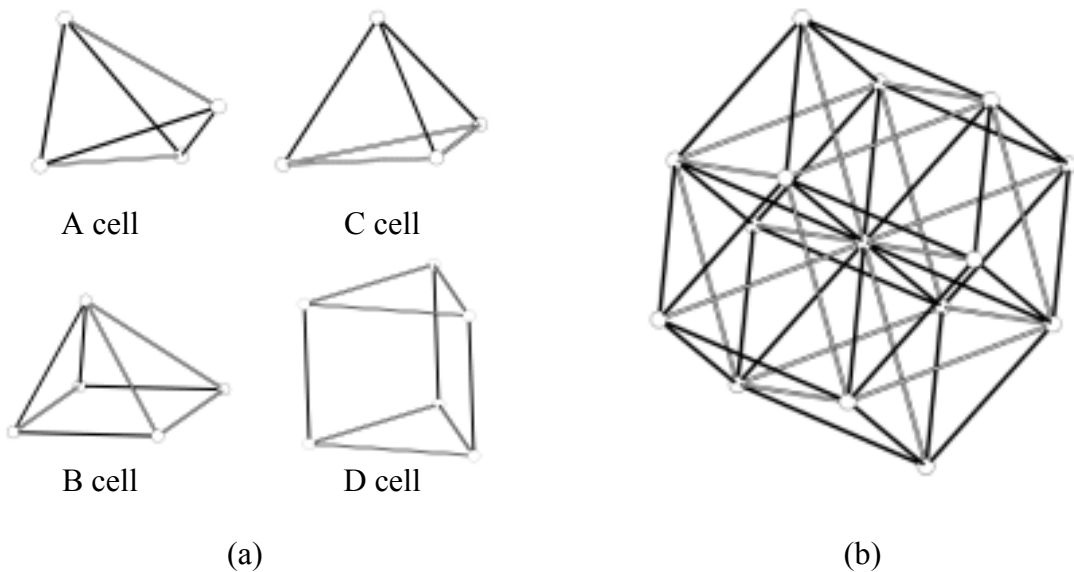

**Fig. 7** Cluster connecting scheme: (a) A, B, C and D cell of the CCT [14], b-bonds ($b$ = 14Å): open style, c-bonds ($c$ = 12Å): filled (black) style; (b) only A cells occur in 1/1-approximants – the nodes of the tiling (and therefore the clusters in our "1/1" model) are packed like atoms in a *bcc* structure.

*Holmium substructure*

As proven by the refinements described in Sec. 3, Ho atoms are located in the second shell of the cluster. Eight Ho atoms form the vertices of a cube of edge length $a$ = 5.4Å. Thus, there are no direct Ho-Ho contacts *intra* cluster. It is exactly the same arrangement as found for the *fci* phase. There is one *inter* cluster contact $d$(Ho1-Ho1) = 2.85Å per Ho atom in the present model of the *si* phase. We have shown for the "2/1" model of the *fci* phase that *inter* cluster direct contacts can be avoided by tilting the cubes with respect to each other in neighbouring clusters [13].
That way two points are achieved: (i) Ho in CN16 (FK polyhedron P) is only surrounded by 12 smaller Zn atoms and 4 slightly smaller Mg atoms while in other comparable crystal structures (*e.g.* HoZn$_3$, oP16, *Pnma* [42]) CN17 is realized for the Ho atoms. The *i*-Mg-Zn-*RE* structures do not seem to offer this coordination number. Steric hindrance then has to be avoided by the absence of Ho ligands in the P coordination shell of a central Ho atom. (ii) Only a superposition of the five possible orientations of a cube inscribed in a pentagon dodecahedron (Fig. 8) generates icosahedral symmetry. Since the title compound exhibits diffraction symmetry *m*-3-5 [15] and *not m*-3 (as our model structure does), *si*-Ho$_{11}$Mg$_{15}$Zn$_{74}$ is assumed containing an ensemble of Ho$_8$ cubes, tilted with respect to each other in adjacent clusters.
In the *fci* phase some Ho is found on glue atom positions. This would correspond to the δ site of Zn5 in the present "1/1"-Ho$_{9.9}$Mg$_{14.8}$Zn$_{75.3}$ model for the *si* phase. However, the temperature factor of Zn5 does not point to this possibility. But as our model is a limited cutout from the real quasicrystal ($r_{max}$ = 17Å – the δ position will split in higher approximant models) we cannot respond clearly to this question. However, assuming that only 5% of Zn5 were replaced by Ho, the measured density would be matched exactly, leaving the overall composition still in its WDX uncertainty range of ±1at%.

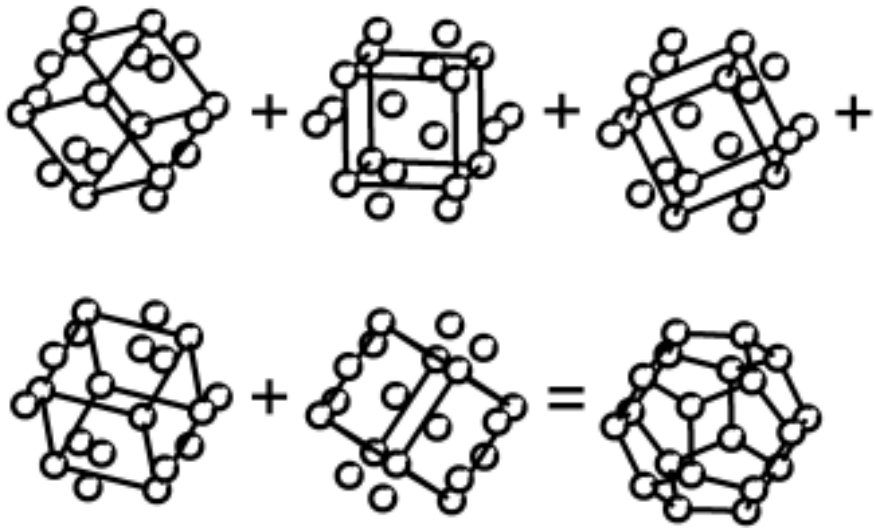

**Fig. 8** 20 atoms form the vertices of a pentagon dodecahedron: a subset of 8 atoms can be combined to the vertices of a cube. The cube edge length is τ times the dodecahedron edge length. Five different orientations of the cube are possible. In *si*-$Ho_{11}Mg_{15}Zn_{74}$ such a shell ($Ho_8Mg_{12}$) is present – the superposition of its five possible orientations is presumed to allow for icosahedral diffraction symmetry.

*Limits and strengths*

The last question discussed above already points at a principal limit of our model: The exact structure beyond the third shell, like the glue atoms and the right cluster connection scheme is out of reach ($r_{max}$ = 17Å). Strictly speaking, we cannot rule out a fourth cluster shell – though the single crystal diffraction photographs [15] point at an architecture close to that of the *fci* phase that contains only 3-shell clusters.

Also, the derived geometries are a somewhat average picture of the real quasicrystal. As a strength, on the other hand, the assignment of different sorts of metal atoms to the different sites *intra* cluster is unambiguous. Their crystal chemical environments turn out as realistic as in periodic crystalline alloys.

However, it can be assumed that the local model can be found in a certain volume anywhere in the quasicrystal because it fits well the quasicrystals PDF data.

Tackling a quasicrystal structure with use of our *local* method, it is a strength where no long range order has to be formulated – at the same time it is an intrinsic drawback that the models are not able to describe the quasiperiodic *long-range* order. Thus the nature of the centering in 6D (*fci versus si*) cannot yet be understood from the (small) local 3D structure elements.

On the other hand again, a strength is the *inductive* character of our method: larger approximant models that are to come in the future will comprise the meaning of this centering directly in 3D space.

*si*-Ho$_{11}$Mg$_{15}$Zn$_{74}$ vs. *fci*-Ho$_9$Mg$_{26}$Zn$_{65}$ [13]

Table 5 highlights the main communities and distinctions of *si* and *fci* Ho-Mg-Zn phases. The main distinction araises from the various Mg/Zn ratios in the phases (0.2 *vs.* 0.4) at virtually constant *RE* content. Lattice parameters, density, mean atomic volume and local distortions are influenced accordingly. Larger Zn content causes occupation of the cluster center in the *si* phase and consequently enlarges the first and second shells (compare Ho$_8$ cube edge lengths). The third shell/glue is made up of Zn atoms only, in total leading to a smaller average volume <*V*> per atom compared to the *fci* phase.

The common basic feature of both phases is the core of the Bergman cluster given by its first two shells. This 45-atom structural element is also known as "Mini-Bergman cluster" or "Pauling triacontahedron" [20]. The center may or may not be occupied: Interestingly, this is exactly the same common structural subunit for various Mg-(Zn,*B*) (*B* = Al or Ga) periodic structures that are known today [20, 22]. The inner part of the Pauling triacontahedron (33-atom cluster "*B*" [43]) is described to make up 80% of *i*-Al-Cu-Fe. This was derived from 6D diffraction data [43].

| structural feature | *si*-Ho$_{11}$Mg$_{15}$Zn$_{74}$ (present work) | *fci* -Ho$_9$Mg$_{26}$Zn$_{65}$ [13] |
|---|---|---|
| Laue symmetry | *m*-3-5 | *m*-3-5 |
| 6D Bravais type | *P* | *F* |
| $a$(6D)/Å | 5.14 [15] | 2 × 5.18 |
| $a_{1/1}$'(3D)/Å | 14.09 | 14.22 |
| basic building block | 105-atom Bergman cluster | 104-atom Bergman cluster |
| center | Zn | void |
| 1st shell | 12 Zn | 12 Zn |
| 2nd shell | 8 Ho + 12 Mg + 12 Zn | 8 Ho + 12 Mg + 12 Zn |
| $a$(Ho$_8$ cube) | 5.4 | 5.3 |
| 3rd shell | 48 Zn + 12 Mg | 60 Zn |
| glue atoms | 1.0 Zn (?) | 0.74 Mg + 0.26 Ho |
| local distortions | like Mg$_2$Zn$_{11}$, *no* FK type | like MgZn$_2$, almost FK type |
| $\rho_{X-ray}$ ($\rho_{meas}$) /gcm$^{-3}$ | 6.46 (6.65) | 6.08 (5.82) |
| <*V*> per atom/Å$^3$ | 17.25 | 17.96 |

**Tab. 5** Comparison of "1/1" hypothetical approximant local models for *si* and *fci* Ho-Mg-Zn phases, basic communities are shaded in grey.

## 5    Conclusion

We explain our new strategy for quasicrystal structure determination: It is to refine local structural models that are hypothetical rational approximants of the quasicrystal. The atomic pair distribution function (PDF) of the quasicrystal is used for the refinements. The PDF has previously to be derived from measured X-ray powder diffraction data.

Application to *si*-Ho$_{11}$Mg$_{15}$Zn$_{74}$ [15] results in a good 3D real space atomic model. The elemental composition of this "1/1"-Ho$_{9.9}$Mg$_{14.8}$Zn$_{75.3}$ model and its density match the experimental values.

The contained atoms are credibly coordinated at credible interatomic distances. The local coordination polyhedra are related to FK coordination shells, though *si*-Ho$_{10}$Mg$_{14}$Zn$_{76}$ does not represent a FK phase in the strict sense. They recall those present in the binary phase $Mg_2Zn_{11}$ (cP39, $Pm$-3; 0/1-approximant) of comparable high Zn content. Thus, atoms in *si*-$Ho_{10}Mg_{14}Zn_{76}$ do not behave as if they were a packing of hard spheres of slightly different radii. The latter is (almost) true for *fci*-$Ho_9Mg_{26}Zn_{65}$ [13].

The structure is made up of 105-atom 3-shell Bergman clusters ($d \approx 15$Å) and some glue atoms (*i.e.* 7% of all atoms in the present model). In contrast to the *fci* phase, the cluster center is occupied by a Zn atom and also in the 3rd shell *only* Zn atoms build a (distorted) soccer ball. Analogous to the *fci* phase, the 1st shell is a $Zn_{12}$-icosahedron, the 2nd shell consists of a $Ho_8$-cube, completed by 12 Mg atoms to a pentagon dodecahedron which in turn is amended by 12 Zn atoms capping the pentagonal faces (45-atom Pauling triacontahedron).

We now can deduce the Pauling triacontahedron including the $RE_8$ cube with a cube edge length of about 5.4Å as a generic feature for *i*-Mg-Zn-*RE* quasicrystals (*RE* = Er or Ho; both *fci* and *si* type). Meeting certain steric conditions, the $RE_8$ cubes are tilted with respect to each other in adjacent clusters and thereby enable diffraction symmetry *m*-3-5.

The clusters are linked sharing hexagonal faces and common edges of the 3rd shell – though the exact cluster connecting scheme cannot yet be made accessible unambiguously *via* our local models. The same holds for the 3D nature of the *P vs. F* type 6D Bravais lattices. Probably a decoration of a (higher approximant or quasiperiodic) CCT model will respond to these still open questions. We further believe that a complementation by other methods, *e.g.* 6D analysis of single crystal diffraction data, will promote a final solution.

### Acknowledgements

We would like to give our sincere thanks to Inga Vasilieva, Boreskov Institute for Catalysis (Novosibirsk) for the ICP elemental analysis of our samples. The Deutsche Forschungsgemeinschaft is acknowledged for the financial support (Schwerpunktprogramm 1031 „Quasikristalle").